\documentclass[twocolumn,preprintnumbers,floatfix,prl,superscriptaddress]{revtex4}
\usepackage{graphicx}
\usepackage{graphicx}
\usepackage{dcolumn} 
\usepackage{bm}
\usepackage{epsfig}
\begin{document} 

\title{Quantum Confinement Induced Molecular Mott Insulating State in La$_4$Ni$_3$O$_8$}

\author{Victor Pardo}
 \email{victor.pardo@usc.es}
\affiliation{Department of Physics,
  University of California, Davis, CA 95616
}
\affiliation{
Departamento de F\'{\i}sica Aplicada, Universidade
de Santiago de Compostela, E-15782 Santiago de Compostela,
Spain
}

\author{Warren E. Pickett}
 \email{wepickett@ucdavis.edu}
\affiliation{Department of Physics,
  University of California, Davis, CA 95616
}


\begin{abstract}
The recently synthesized layered nickelate La$_4$Ni$_3$O$_8$, with its cuprate-like NiO$_2$ layers,
seemingly requires a Ni1 ($d^8$)+2Ni2 ($d^9$)
charge order, together with strong correlation effects, to account for its insulating behavior. Using
density functional methods including strong intra-atomic repulsion (Hubbard U), we obtain an insulating
state via a new mechanism: {\it without charge order}, Mott insulating behavior arises
based on quantum coupled, spin-aligned 
Ni2-Ni1-Ni2 $d_{z^2}$ states across the trilayer (rather than based on atomic states), with
antiferromagnetic ordering within layers. The weak and frustrated magnetic coupling between cells
may account for the small spin entropy that is removed at the N\'eel transition at 105 K and the lack
of any diffraction peak at the N\'eel point.
\end{abstract}

\maketitle

Nearly 25 years since the discovery of high temperature superconductivity (HTS), the question persists: why
does HTS occur only in copper oxides and not other layered oxides.  Nickelates
with cuprate-like structures\cite{review_pickett} have seemed to be the most likely candidates: the Cu$^{2+}$ 
$d^9$ configuration with active $d_{x^2-y^2}$ orbital maps onto the Ni$^{1+}$ configuration, but this is known to be a
problematic charge state. Nevertheless, difficult charge states may sometimes be stabilized, such as the
nominal Ni$^{1+}$ (``infinite layer'') compound LaNiO$_2$ prepared by topotactic 
exchange of oxygen\cite{kawai_2009}.
Cuprate-like electronic structures in other oxides have been pursued since soon after HTS were discovered.
The charge conjugate analog Sr$_2$VO$_4$
($d^1$ vs. $d^9$) presented hope\cite{sr2vo4_expt,sr2vo4_bands} and displays magnetic and 
orbital ordering\cite{sr2vo4_data}
but has never shown superconductivity.
Layered nickelates have been seen as the best mimic of HTS if the $d^9$ Ni$^{+}$ charge state can be
stabilized, and theoretical interpretation of the few known prospects has begun\cite{anisimov_1999,kwlwep}.
The LaNiO$_3$/LaAlO$_3$ superlattice\cite{giniyat,lanio_khaliullin} is an example that has been
suggested to produce an HTS-like Fermi surface, but is based on the more common Ni$^{3+}$ ion and
has not yet produced superconductivity.

The discovery by Greenblatt's group of a Ruddleston-Popper sequence\cite{lanio_2006} of phases La$_{n+1}$Ni$_n$O$_{2n+2}$ with $n$ cuprate-like
NiO$_2$ layers has reinvigorated interest in nickelates, with a focus on La$_4$Ni$_3$O$_8$ (``La438'') where
stoichiometry is attained\cite{lanio_2007,lanio_polt1}. In an ionic picture, the mean Ni valence in 
La$_{n+1}$Ni$_n$O$_{2n+2}$ is $+(n+1)/n$, {\it i.e.} metallic.
However, La438 is highly insulating\cite{lanio_curro_1},
with a room temperature resistivity $\sim$ 10$^3$ $\Omega$-cm increasing by six orders of magnitude
down to 25 K. Magnetic ordering is observed\cite{lanio_curro_1} at T$_N$=105 K,
with NMR data indicating an antiferromagnetic (AFM) transition\cite{lanio_curro_2}.  The entropy loss
at T$_N$ is only $\frac{1}{3}$R$\ln{2}$ per Ni, suggesting unconventional magnetic behavior in this system.

\begin{figure}[ht]
\begin{center}
\includegraphics[width=0.80\columnwidth,draft=false]{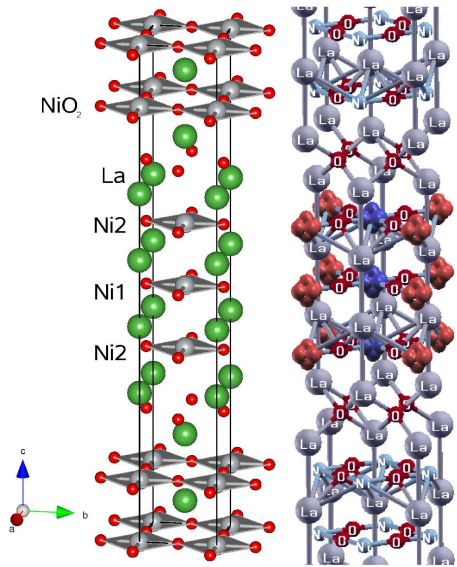}
\caption{Left side:
structure of La$_4$Ni$_3$O$_8$, showing the three NiO$_2$ ``infinite layer''
structures sandwiched on either side by the fluorite La/O$_2$/La blocking layer.
The Ni1 and Ni2 designations of the NiO$_2$ trilayer are provided. Right side:
spin density isocontour plot of the ground state of  La$_4$Ni$_3$O$_8$ 
from LDA+U calculations for the $\sqrt{2}\times \sqrt{2}\times 1$ cell. The 
different colors denote the two different spin directions. Magnetic order is AFM within layers, with spins
aligned along the perpendicular (vertical) axis.}
\label{spindens}
\end{center}
\end{figure}

La438 crystallizes in a tetragonal unit cell, with space group I4/mmm (\#139) and lattice parameters $a$=3.9633 \AA\,
$c$=26.0373 \AA. The structure, pictured in Fig. \ref{spindens}, consists of three NiO$_2$ ``infinite 
layer'' planes  separated by layers of La$^{3+}$
cations but no oxygen.  On either side of this trilayer slab lies a fluorite La/O$_2$/La layer, with the same structure
as the corresponding layer in LaFeAsO.  While in a sense isoelectronic with doped cuprates as it seems to
involve some mix of Ni$^{1+}$ and Ni$^{2+}$ cations, the electronic and magnetic structure of
the Ni ions is still unclear.  As mentioned, the low oxidation state Ni$^{1+}$ is enigmatic;
to obtain such a configuration for Ni is not easy, and
these Ruddleston-Popper phases are unstable at moderate ($\sim$375 $^{\circ}$C) temperature\cite{lanio_2006} to decomposition
into La$_2$O$_3$ and Ni metal.  In the Ni$^{1+}$ ``infinite-layer'' 
compound LaNiO$_2$ it was found to
be impossible\cite{kwlwep} to produce a Mott insulating state as might be anticipated; in fact the compound is metallic
at least roughly as calculated.  La438 consists of three layers of LaNiO$_2$ subject to quantum confinement,
as we demonstrate below.  

In La438 a simple ionic picture will give Ni configurations $d^8$ + 2$d^9$, which would give the natural 
identification of the one `inner' Ni1 $\equiv$ Ni$^{2+}$($d^8$) and two outer Ni2 $\equiv$ Ni$^{1+}$($d^9$)
cations, as labeled in Fig. \ref{spindens}.
Ni1-Ni2 layers are separated by only 3.25 \AA, much less than the in-plane separation $a$=3.93 \AA, a
characteristic that will become important below.
The Ni trilayers are separated by the La/O$_2$/La unit, and an ($a/2,a/2$,0) translation between successive
NiO$_2$ trilayers reinforces very small
interplanar coupling between cells, which will be shown to produce a natural quantum confinement within the trilayers. 


Here we study the electronic structure of the compound by {\it ab initio}
techniques to understand the electronic and magnetic structure of La438 and to identify
the differences and similarities with respect to the cuprates.
Our electronic structure calculations were  performed within density functional
theory \cite{dft_2} using the all-electron, full potential code {\sc wien2k} \cite{wien}
based on the augmented plane wave plus local orbital (APW+lo) basis set \cite{sjo},
and the experimental atomic positions\cite{lanio_curro_1}.
The generalized gradient approximation\cite{gga} was used.
To deal with  strong correlation effects we apply the LDA+U
scheme \cite{sic1,sic2} including an on-site repulsion U and Hund's coupling J 
for the Ni $3d$ states.  Results were not very dependent on the specific values of U and J
in the reasonable range, and we
report results with U = 4.75 eV, J = 0.68 eV, values very similar to those determined from
constrained density functional calculations\cite{lanio_curro_1}.

The bands within LDA (U=0) are metallic as expected, and in fact the bands crossing the Fermi energy E$_F$ are
dispersive (bandwidth of $\sim$2-3 eV).  These bands do not suggest substantial effects of 
strong correlation, and Poltavets {\it et al.} suggested\cite{lanio_curro_1} Fermi surface driven spin
density wave (SDW) order, which might destroy part of the Fermi surface and account for the magnetic transition.
The earlier calculations, even those using LDA+U, produced only metallic states\cite{lanio_curro_1}.
The key to obtaining a lower energy, and insulating, state is to allow extra freedom by doubling the
lateral cell of La438, and we found that in-plane AFM order is readily obtained. Various configurations and types 
of spin order were tried, but {\it no inter-plane AFM} order could be obtained; moments along the
$z$ axis always want to align within the trilayer.

We discuss only this most energetically stable state, comprised of AFM-ordered Ni layers with spins aligned
along the $z$ axis. The spin density, pictured in Fig. \ref{spindens}, makes evident both the type of spin
ordering and the character (shape) of the spin density, which reveals both $d_{x^2-y^2}$ 
and $d_{z^2}$ contributions. There is indeed some charge difference 
with muffin-tin charges of 8.52 (Ni1) and 8.64 (Ni2), and different moments of 1.4 and 1.3 $\mu_B$ respectively.
It is understood that different formal charge states may involve surprisingly reduced difference in
actual charge; it is magnetic states that are more instructive.  While the moments we calculate might be 
consistent with an S=1 configuration ($d^8$) reduced by hybridization, neither moment (1.3-1.4 $\mu_B$) can be ascribed
to an S=$\frac{1}{2}$ $d^9$ ion (which would have a maximum moment of 1 $\mu_B$). Without $d^9$ ions, the
ionic, charge-ordered picture becomes untenable, and a new mechanism for producing the gap must operate.

\begin{figure*}[ht]
\begin{center}
\includegraphics[width=0.3\textwidth,draft=false]{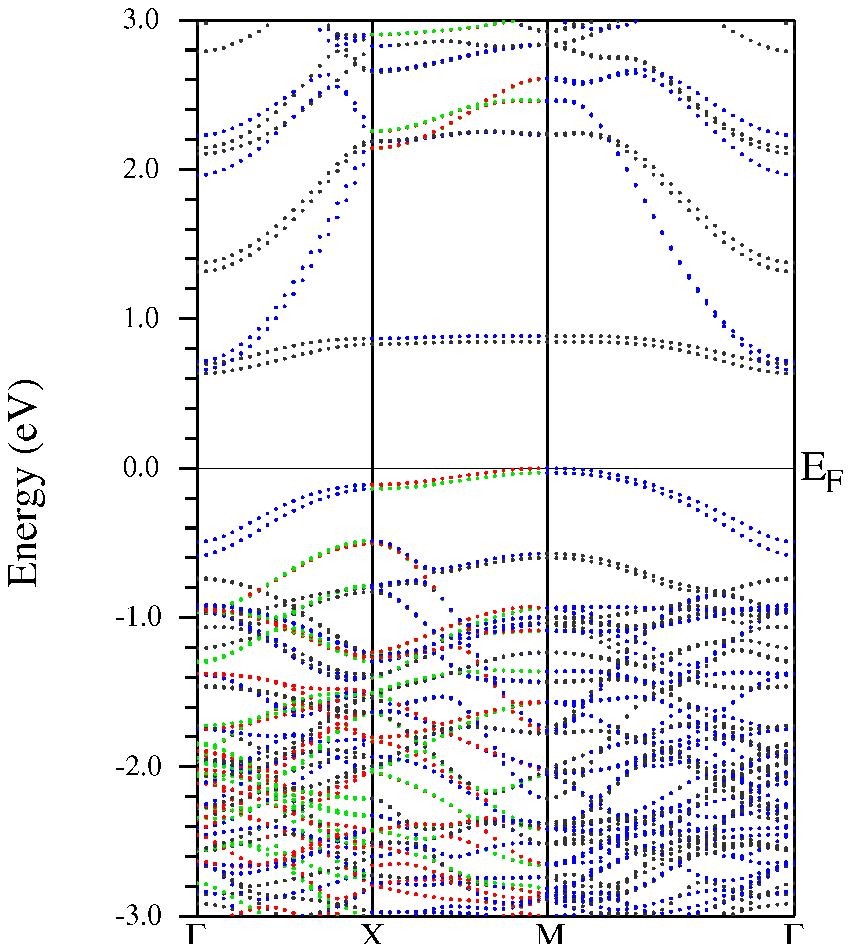}
\includegraphics[width=0.3\textwidth,draft=false]{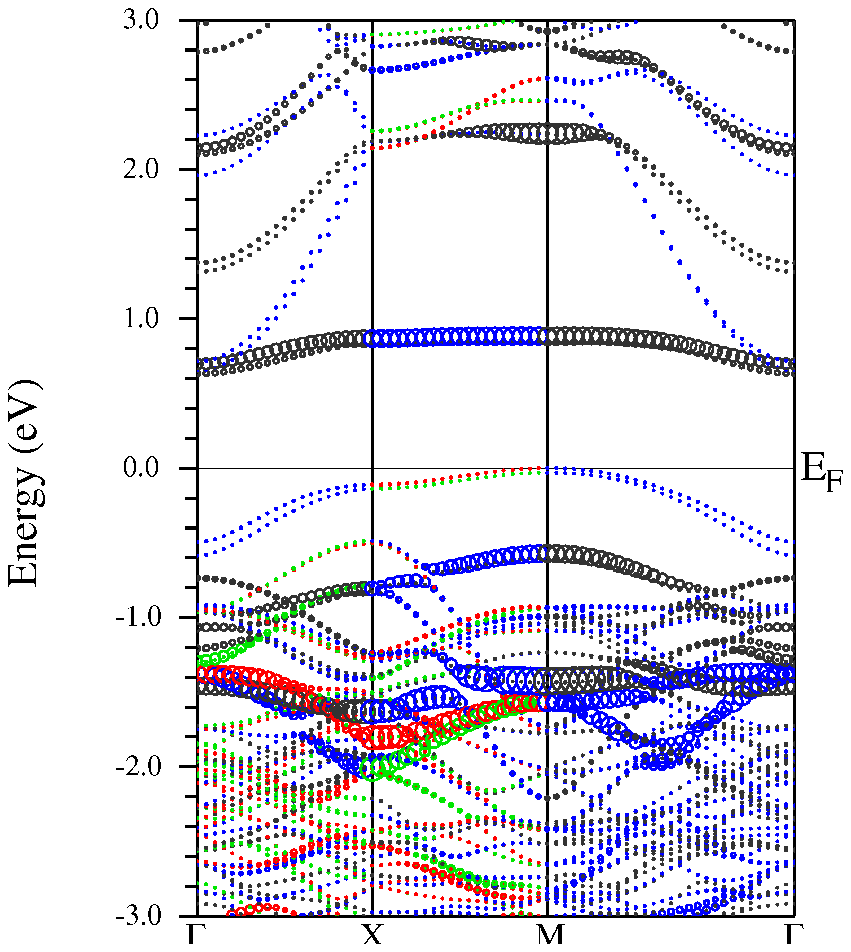}
\includegraphics[width=0.3\textwidth,draft=false]{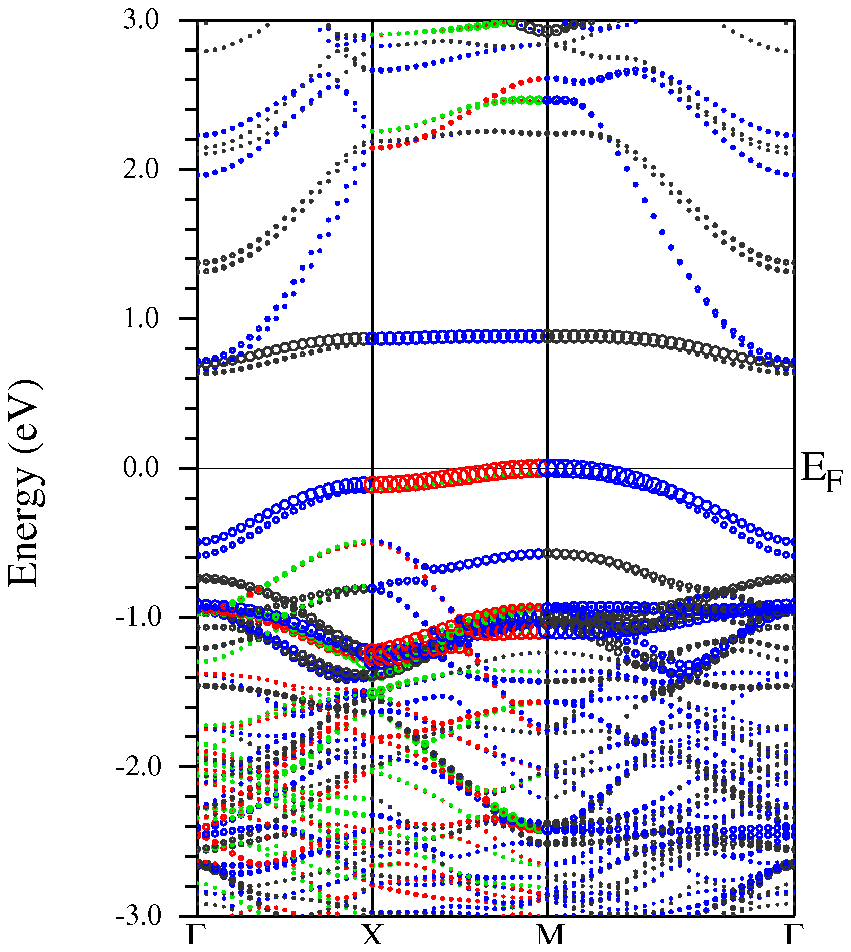}
\caption{The Mott insulating band structure of La$_4$Ni$_3$O$_8$ for the calculated ground state
of Fig. \ref{spindens}. Left panel: the basic plot, allowing nearly degenerate bands to be distinguished.
Other panels: La$_4$Ni$_3$O$_8$ fatband representation emphasizing (middle panel) Ni2, and (right
panel) Ni1, $d$ character. Note that
the band at the bottom of the gap has vanishing Ni1 character.
}\label{bands}
\end{center}
\end{figure*}

Our LDA+U band structure, shown in Fig. \ref{bands}, produces a Mott insulating state, but of an unusual type that
requires broadening one's view of how these correlation effects operate.
In the AFM state, the spin-up and -down band structures are of course identical. Since
the valences of La$^{3+}$ and O$^{2-}$ are clear, the Ni ions are on average Ni$^{1.33+}$, with charge $d^{8.67}$
making the maximum moment around 1.33 $\mu_B$, which is in fact what they are in our ground state. The Ni ions are
thus fully polarized, meaning the majority orbitals are completely occupied and removed from consideration.  Also, the
minority $d_{x^2-y^2}$ orbitals are empty, centered at +3.5-4 eV on the energy scale of Fig. \ref{bands}. Thus
the $d_{x^2-y^2}$ orbital contributes one unit to the hole count, 1 $\mu_B$ to each Ni moment, and the
in-plane superexchange coupling, but otherwise
is of no interest.  The relevant orbitals are solely the partially occupied minority $d_{z^2}$ orbitals 
on the six Ni ions in
the AFM cell, which must accommodate one hole per triple (two in the AFM cell).  The central question then is:
how can a Mott insulating state, as we obtain, arise at ``2/3 filling''
if charge disproportionation into 2 Ni$^{1+}$ +  Ni$^{2+}$ does not occur?

\begin{figure*}[ht]
\begin{center}
\includegraphics[width=0.9\textwidth,draft=false]{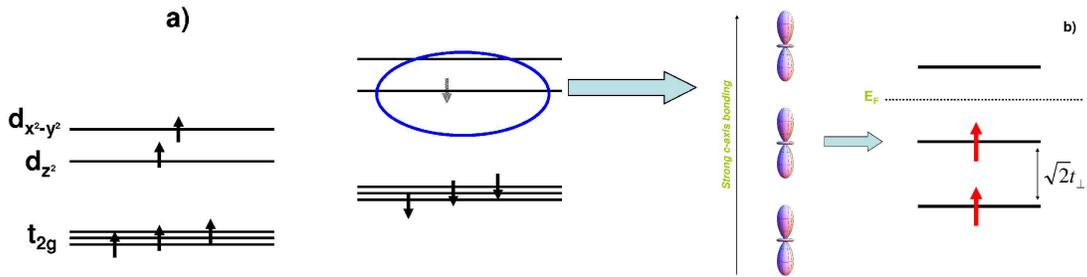}
\caption{Energy level diagram. The intraatomic levels are shown on the left, with occupied
majority states on the far left, partially occupied minority levels just to the right.
The right panel indicates how the $d_{z^2}$ orbitals on the Ni2-Ni1-Ni2 triple couple to give
a gap at 2/3 band filling, following Eq. \ref{equation}.
}\label{energylevels}
\end{center}
\end{figure*}

The in-plane AFM order, as is common, drastically reduces bandwidths (spin-conserving hopping takes place only between second
Ni neighbors in the plane), so taking account of
Ni $d$ interplanar coupling $t_{\perp}$ (between Ni1 and Ni2) becomes important.  For each Ni2-Ni1-Ni2 triple 
along the $z$ axis with neighbors
coupled by $t_{\perp}$ (with zero site energy, neglecting the small on-site energy difference), the eigenvalues and eigenvectors are
\begin{eqnarray}
   E_j &=& 0, ~~ \pm \sqrt{2}t_{\perp}, \nonumber \\
  |E_j>       &=& \frac{1}{\sqrt{2}}(1,0,-1), ~~ \frac{1}{2}(1,\pm \sqrt{2},1).
\label{equation}
\end{eqnarray}
The non-bonding (odd symmetry) state at $E_{j=0}$=0 which does not involve the Ni1 site is flanked below
and above by even symmetry bonding ($E_{+}$) and antibonding ($E_{-}$) combinations. This coupling is pictured at the
right of Fig. \ref{energylevels}

In the middle and right panels of Fig. \ref{bands} the Ni $d$ characters are emphasized for Ni2 and Ni1, respectively.  The bands display both the
spectrum (most clearly evident near the M point) and the (Ni1 vs. Ni2) character of this ``triatomic molecule''.
For example, the band at the bottom of the gap is (non-bonding)
purely Ni2, which is $E_0$, while the other two bands have both Ni1 and Ni2 character ($E_{+}, E_{-}$). Two of these levels will be filled (below the Fermi
level) and one empty (above the Fermi level). The gap size, which is 0.6 eV here, will depend both on the coupling
$t_{\perp}$ and the value of U.

Since previous LDA+U studies produced only a metallic result, one can understand the electronic structure 
only using a quantum-coupled unit, a {\it molecular trimer basis}, for the trilayer,
from which on-site repulsion produces a (Mott) insulating state in agreement with observation\cite{lanio_curro_1}.
The intralayer AFM magnetic coupling is strong and is
mediated by d$_{x^2-y^2}$ superexchange, as in cuprates. Between NiO$_2$ layers, the coupling is FM direct exchange
arising from the $d_{z^2}$ orbitals (Wannier functions). This FM coupling is robust in the sense that
we could not obtain antialigned
layer magnetic order.  Experimental data suggest an in-plane magnetic 
coupling with J$_{\parallel}\approx$ 120 K\cite{lanio_curro_1}, which is substantial but
an order of magnitude smaller than in cuprates.

The trilayer-to-trilayer coupling will be superexchange through the fluorite La/O$_2$/La spacer layer that can be seen
in Fig. \ref{spindens}.  The $k_z$ bandwidth and related
dispersion can be estimated from the splitting of the band below the gap near -0.5 eV at $\Gamma$ (this splitting
vanishes at $\pi/c$).  The hopping $t_z \sim 15$ meV from a Ni2 ion to four Ni2 ions across the 
blocking layer leads to $J_z$ = 4$t_z^2/U \sim$ 0.2 meV $\sim$ 2 K.  
Such small interlayer coupling enters T$_N$ logarithmically and can still give rise to magnetic order. 
However, the interslab coupling is frustrated as well as being small. These facts suggest that some of the
entropy will not be lost at T$_N$=105 K (full magnetic order along the $z$ axis is not established), 
consistent with the small observed entropy change and also with the absence of a new diffraction 
peak\cite{lanio_curro_1} below T$_N$.


The state we have obtained and analyzed has insulating, AFM order in-plane in common with the HTS cuprates, as well as the 
obvious similarities in structure. The differences are however substantial.  First, excellent insulating character
and relatively strong superexchange-mediated AFM order is obtained {\it without} integer formal oxidation states on the
Ni ions. Such behavior has not been reported previously for transition metal oxides. 
Second, the $d_{x^2-y^2}$ hole on each Ni ion is robust, and doped carriers go only into $d_{z^2}$ states; hence
the AFM order (driven by $d_{x^2-y^2}$ coupling) may survive to much larger doping level 
than in cuprates. Orbital order is ruled out. 
Third, holes will enter the non-bonding band that
is confined to the outer Ni2 layers, so beyond the insulator-to-metal transition these two conducting layers
will be separated by a non-participating Ni1 middle layer; the conductivity will be both orbital and layer
specific.  Doped electrons will go into all three Ni layers, but with half of the charge on the central Ni1 layer.

The insulator-to-metal transition in either case will likely occur within the AFM ordered phase rather than 
occurring simultaneously. Superconductivity
coexisting with AFM order is known to occur, but the competition tends to be very unfavorable for superconductivity. 
Moreover, the magnetic interactions between the doped carriers (in $d_{z^2}$ states) will be very weak, and very
unlike cuprates where carriers are doped into the $d_{x^2-y^2}$ states.  Though substantially different from 
cuprates, doped La438 should be an extremely
interesting system, and doping should be pursued.

We acknowledge informative communication with N. J. Curro.
This project was supported by DOE grant DE-FG02-04ER46111.


\end{document}